\input{aipcheck}

\documentclass[
    ,final            
  ]
  {aipproc}

\layoutstyle{6x9}

\begin{document}

\title{Generalized Parton Distributions\\ 
		and\\
 		Deep Exclusive Reactions:\\
		Present Program at JLab}

\author{Michel Gar\c con}{
  address={DAPNIA/SPhN, CEA-Saclay, 91191 Gif-sur-Yvette, France}
}

\begin{abstract}
 We review briefly the physical concept of generalized parton
 distributions and the experimental challenges associated 
 with the corresponding measurements of deep exclusive reactions.
 The first data obtained at Jefferson Lab for exclusive photon (DVCS) 
 and vector meson (DVMP) electroproduction
 above the resonance-excitation region are described.
 Two upcoming dedicated DVCS experiments are presented in some detail. 
\end{abstract}

\maketitle


\section{Generalized Parton Distributions}

The physical concepts and underlying theory of generalized parton
distributions (GPD) were to be reviewed at this Workshop by A. Sch\"afer,
who could not attend.
We certainly missed his insight and rigor on the subject. 
In a much more heuristic way than he would have done, I recall some
of the main features of GPDs.

In the Bjorken limit, exclusive processes of the type $\gamma^*p\to
p\gamma,\pi,\rho,\omega\cdots$ are described in term of a handbag
diagram. The corresponding amplitude factorizes~\cite{Ji97,Col97} 
into a hard scattering
process involving one participant parton (e.g. $\gamma^*q\to\gamma q$)
and a soft rearrangement in the nucleon parameterized by four GPDs,
denoted by $H,E$ (``unpolarized'' or more exactly independent of the
quark helicity) and $\tilde{H},\tilde{E}$ (``polarized''). 
In addition to the scale variable $Q^2$, these are
functions of three variables:
$x$, the average longitudinal  parton momentum fraction before and after
the hard scattering; $\xi$, half the difference between these two momentum
fractions, and a four-momentum transfer squared $t$.

An oversimplified physical image can be
given in terms of ordinary quantum mechanics. 
Schematically, a nucleon reduced to a 3-quark configuration
can be described by a wave function 
$\psi(x_1,\vec{k}_1, x_2,\vec{k}_2,x_3,\vec{k}_3)$ where $x$ is
again a parton longitudinal momentum fraction and $\vec{k}$ its transverse
momentum. An ``ordinary'' parton distribution is given by the
probability integral
\begin{equation}
q(x)\sim\int |\psi(x,\vec{k}_1, x_2,\vec{k}_2,x_3,\vec{k}_3)|^2[dX]
\end{equation}
with $[dX] = \delta(x+x_2+x_3-1)\delta^{(2)}(\vec{k}_1+\vec{k}_2+\vec{k}_3)
		dx_2dx_3d\vec{k}_1d\vec{k}_2d\vec{k}_3$.
In this representation, a GPD can be thought of as an overlap integral
\begin{equation}
H(x,\xi,t)\sim\int 
  \psi^*(x-\xi,\vec{k}_1+\vec{\Delta}_{\perp}, x_2,\vec{k}_2,x_3,\vec{k}_3)
  \cdot\psi(x+\xi,\vec{k}_1, x_2,\vec{k}_2,x_3,\vec{k}_3) [dX]
\end{equation}
The off-diagonal character of the GPD appears in the change of longitudinal
momentum: a GPD thus expresses the {\it coherence} between states of
different longitudinal momenta. 
Kinematically, this rearrangement cannot
occur without a non-zero squared momentum transfer $t=f(\xi,\vec{\Delta}_{\perp})$.
Through a Fourier transform with respect to $\vec{\Delta}_{\perp}$ at
$\xi=0$, GPDs can be related to spatial transverse distributions of
partons. Thus, through the $x$-$t$ correlation, the GPDs contain information
about the transverse distribution of partons of given $x$. This opens the
way for a femto-photography of the nucleon. Qualitatively, this correlation
between the transverse position $\vec{b}$ and the longitudinal momentum
$\vec{p}$ gives also access to information about the orbital angular
momentum of the partons ($\vec{b}\wedge\vec{p}$).

This picture can be extended to include quark-antiquark configurations
in the initial and/or final states.

\section{Experimental challenges}
\begin{figure}
  \label{fig:res}
  \includegraphics[height=.5\textheight]{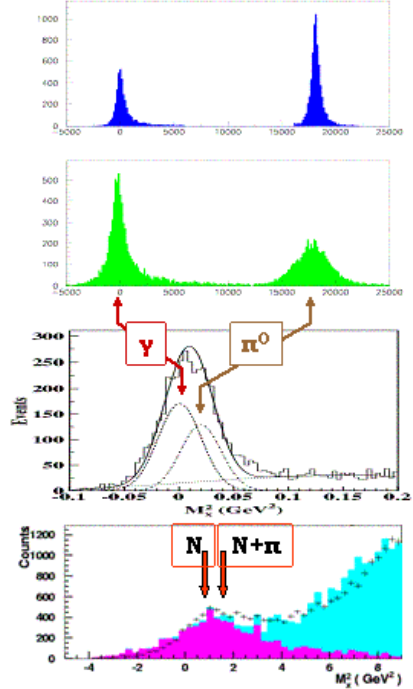}
  \caption{Squared missing mass for performed VCS or DVCS experiments. From top
  		to bottom:
  		$ep\to epX$ from MAMI (850 MeV beam energy)~\cite{Fon02},
  		from JLab/Hall A (4 GeV) ~\cite{Fon02},
  		and from JLab/CLAS (4.2 GeV)~\cite{Ste01},
  		$ep\to e\gamma X$ from HERMES (27 GeV)~\cite{Air01}.
	  }
\end{figure}

At high virtuality $Q^2$ of the exchanged photon, 
the measurements of exclusive production reactions 
are very challenging. 
Precision measurements require high luminosity $\times$ acceptance
since the cross sections are small. Exclusivity is necessary for
a quantitative comparison with theory and for a detailed study of 
scaling laws. This puts high demands on resolution and/or
complete coverage of reaction products. For example, in the
$ep\to ep\gamma$ process, it is in principle sufficient to
identify two of the three particles in the final state and
measure their momentum vector (this is the case in the first
two graphs of figure~\ref{fig:res}). In practice, as beam and secondary
particles energies increase, it becomes increasingly difficult
to separate DVCS from exclusive $\pi^0$ production when measuring
only the scattered electron and proton (third graph of fig.~\ref{fig:res}). 
It is then
necessary to detect all three particles in the final state.

\section{Deeply Virtual Compton Scattering}

The first measurements of beam spin asymmetry in the $\vec{e}p\to ep\gamma$
exhibit an azimuthal dependence
characteristic of the interference of a helicity conserving process
such as DVCS with the Bethe-Heitler process~\cite{Ste01,Air01}. Moreover
the magnitude of this asymmetry is agreement with models based on GPDs,
including small higher-twist contributions and next-to-leading order
$\alpha_S$ corrections. Note however that the
same results may also be described within a Regge model~\cite{Can03}.

Since then, other CLAS data taken at beam energies of 4.8 and 5.75 GeV
are being analyzed as well. This is still based on the study
of $ep \to epX$ configurations. At 4.8 GeV, the resolution still allows
to perform a shape analysis of the missing mass spectra, similar to
the procedure in Ref.~\cite{Ste01}, thus separating the $\gamma$ and $\pi^0$ 
contributions~\cite{Gav04}. 
An exemple of such a separation is given in the third graph of fig.~\ref{fig:res},
where each contribution is independently calibrated using subsets
of unambiguously identified events~\cite{Ste01}.
At 5.75 GeV, the missing mass resolution worsens
and precludes such an analysis. The pion contribution is then minimized
(but not yet quantified) through an approriate choice of kinematical cuts, 
guided by a Monte-Carlo simulation of both processes~\cite{Ava04}.
In both cases, earlier results are confirmed, and the higher
statistics allow for a finer binning, in $Q^2$, $x_B$ and/or $t$.
The measured beam spin asymmetry, at fixed $x_B$ and $t$, 
is consistent with a $1/Q$ behaviour, as expected from the leading-twist 
(handbag) contribution to DVCS.

Target spin asymmetries, using a polarized NH$_3$ target, were investigated
as well~\cite{Che04}. Here again the contribution of the $ep\to ep\pi^0$
process remains to be quantified.

\begin{figure}
  \label{fig:exp}
  \includegraphics[height=0.4\textheight]{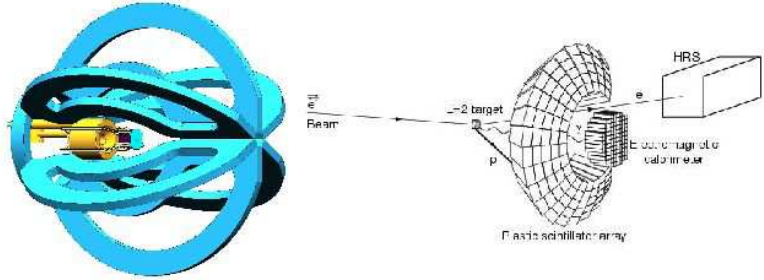}
\vspace{-4cm}
  \caption{Shematic views of dedicated CLAS (left) and Hall A (right) 
  	   DVCS experiments. The standard CLAS detection is not shown;
  	   the new solenoid and calorimeter are positioned at the
  	   center of the CLAS toroidal magnet.
	  }
\end{figure}

Dedicated experiments now aim at detecting the three particles in the final
state, in order to ensure a complete exclusivity in the measurements.
See figure~\ref{fig:exp}. 

In CLAS~\cite{Bur01}, the photons emitted in a forward cone (3-14$^{\circ}$) will
be detected in a new inner calorimeter, consisting of 424 small lead-tungstate
crystals read by avalanche photodiodes. In addition, a specifically designed
two-coil solenoid will focus the background low energy electrons
along the beam axis. A 100-crystal prototype was tested with success
in December 2003. The whole experiment is to be mounted in February 2005.

In Hall A~\cite{Ber00}, a lead-fluoride calorimeter is mounted in direct view of
the target, along the direction of the virtual photon, for each of
three spectrometer settings for the scattered electron. The recoil
protons are detected in a concentric annular plastic scintillator array.
The experiment is currently running at the designed luminosity of
10$^{37}$ cm$^{-2}$s$^{-1}$. A follow-up experiment will study
DVCS on the neutron~\cite{Ber03}. 
The quasi-free $e(n)\to en\gamma$ reaction will
be measured using a liquid deuterium target and complementing the
proton array with thin scintillator for proton/neutron discrimination.
The DVCS beam spin asymmetry on the neutron 
has been shown to be more sensitive to the quark total angular
momentum in the nucleon.

Double DVCS ($ep\to epe^+e^-$ proceeding through 
$\gamma^*q\to\gamma^*q$) has the potential to access
GPDs at a tunable kinematical point $x=f(\xi,q'^2)$,
where $q'^2$ is the virtuality of the outgoing photon
(or equivalently the invariant mass of the corresponding
lepton pair)~\cite{DDVCS}. Positrons were cleanly selected in
CLAS and $ep\to epe^+X$ configurations analyzed~\cite{Mor03}. 
Although
$ep \to ep e^+e^-$ events could be clearly identified,
the analysis faces a theoretical and experimental challenge:
there are two electrons in the final state. Antisymmetrisation
was not taken into account in the calculations. Experimentally,
the kinematical variables depend on the identification
of an electron either as scattered or coming from the lepton pair.
One of the two possible choices result in the observation
of clear peaks associated with very low $Q^2$ vector meson
production. Our preliminary conclusion is then that the observed
$ep\to epe^+e^-$ events cannot be interpreted in the
framework of DDVCS and GPDs.
 
\section{Deeply Virtual Meson Production}

The hard electroproduction of mesons, or DVMP,
should also proceed through the same handbag diagram.
In this case, factorization has been demonstrated for longitudinal
virtual photons~\cite{Col97}. According to the nature of the emitted
meson, one probes different GPDs ($H$ and $E$ for vector mesons,
$\tilde{H}$ and $\tilde{E}$ for pseudoscalar mesons) and
different flavor combinations (e.g. $\rho$ vs $\omega$, or
$\pi^0$ vs $\eta$). The difficulty here is in isolating the
contribution of longitudinal photons and reaching high enough
$Q^2$, since the additional gluon exchange necessary to produce the meson
moves the asymptotic regime to higher values of $Q^2$.

Longitudinal cross sections have been extracted for $\rho$
electroproduction at 4.2 GeV~\cite{Had04}. A similar analysis
for higher energy (5.75 GeV), with higher statistics data,
is underway. In this channel, both the CLAS and HERMES data
are in semi-quantitative agreement with a model based on
GPDs, and including an effective description of higher-twist
effects.

The $\phi$~\cite{San04} and $\omega$~\cite{Mor04} channels 
have been analyzed as well. In the latter case, a determination
of the emitted $\omega$ spin density parameters led to the
conclusion that helicity is not conserved between the virtual
photon and the emitted meson. Thus the handbag diagram does
not dominate the reaction $ep\to ep\omega$ and the longitudinal 
cross sections could not be extracted. 
The $\omega$
channel thus appears the most challenging one to access the GPDs.
Within a Regge approach,
it was shown that this behaviour is due to the dominantly
transverse $t$-channel $\pi^0$ exchange, or rather the exchange
of the corresponding saturating Regge trajectory~\cite{Lag04}. 
Only for the $\omega$ channel is this specific exchange contribution
dominant.
This calculation
gives a new insight in the high-$Q^2$ and high-$t$ behaviour of such
exclusive reactions~\cite{Lag04}.

\section{Conclusions}

Generalized Parton Distributions have merged as a powerful, attractive
and unifying concept for the nucleon structure. The relation
between GPDs and deep exclusive reactions still needs experimental
validation. There are experimental indications that the handbag
diagram is at work in DVCS and this is being studied with increased 
statistics data sets at CLAS. In DVMP, extensive data sets on
vector meson production, also from CLAS, will be coming shortly.
Dedicated DVCS experiments, in Hall A (2004) and CLAS (2005), will measure
beam spin asymmetries, detecting all three particles in the final 
state to ensure full exclusivity.
They should establish on firm grounds the validity
of the approach. Detailed tests of scaling will be performed.
If scaling is indeed observed, or deviations thereof understood,
these experiments, in conjunction with theoretical input, 
will provide the first significant measurements of GPDs.

%

{}

\end{document}